\documentclass[final,3p,times]{elsarticle}
\usepackage{amssymb}
\usepackage[fleqn]{amsmath}
\usepackage{amsthm}
\usepackage{subfigure}
\usepackage{graphicx}

\biboptions{numbers,sort&compress}
\usepackage{hyperref}
\journal{journal}

\begin{document}

\begin{frontmatter}

\title{General mixed multi-soliton solution to the multi-component Maccari system}

\author{Zhong Han$^{{\rm a,b}}$}

\author{Yong Chen$^{{\rm a,b}}$ \corref{cor1} }

\ead{ychen@sei.ecnu.edu.cn}
\cortext[cor1]{Corresponding author.}

\address{$^{{\rm a}}$Shanghai Key Laboratory of Trustworthy Computing, East China Normal University, Shanghai, 200062, People's Republic of China}
\address{$^{{\rm b}}$MOE International Joint Lab of Trustworthy Software, East China Normal University, Shanghai, 200062, People's Republic of China}

\begin{abstract}
Based on the KP hierarchy reduction method, the general bright-dark mixed multi-soliton solution of the multi-component Maccari system is constructed. The multi-component Maccari system considered comprised of multiple (say $M$) short-wave components and one long-wave component with all possible combinations of nonlinearities including all-focusing, all-defocusing and mixed types. We firstly derive the two-bright-one-dark (2-b-1-d) and one-bright-two-dark (1-b-2-d) mixed multi-soliton solutions to the three-component Maccari system in detail.
For the interaction between two solitons, the asymptotic analysis shows that inelastic collision can take place in a $M$-component Maccari system with $M \geq 3$ only if the bright parts of the mixed solitons appear at least in two short-wave components.
The energy-exchanging inelastic collision characterized by an intensity redistribution among the bright parts of the mixed solitons. While the dark parts of the mixed solitons and the solitons in the long-wave component always undergo elastic collision which just accompanied by a position shift.
In the end, we extend the corresponding analysis to the $M$-component Maccari system to obtain its mixed multi-soliton solution. The formula obtained unifies the all-bright, all-dark and mixed multi-soliton solutions.
\end{abstract}

\end{frontmatter}

\section{Introduction}
In the investigation of nonlinear wave dynamics, it is important to extend the study to multi-component counterparts since many complex systems such as nonlinear optical fibres \cite{pre4}, Bose-Einstein condensates \cite{bose} etc usually involve more than one component.
On the other hand, the study of multi-component system is also of great interest as the interaction of multiple waves may result in some new physical processes \cite{pre2,pre3,xiaoen}. Of particular interest is the multi-component generalization of the celebrated nonlinear Schr\"{o}dinger (NLS) equation \cite{gener3,gener4,feng,yan,ling1}. During recent years, the studies on multi-component systems which describe the interaction between short wave packets with long waves have also received many attentions \cite{lak1,lak3,chen1,chen2}.
In the real physical systems, the nonlinearities may be positive or negative, depending on the concert physical situations \cite{segure}. Particularly, in Bose-Einstein condensates \cite{bose}, the nonlinear coefficient is positive or negative if the interaction of the atoms is repulsive or attractive.
It has been shown that in multi-component systems \cite{lak1,lak3,chen2}, the bright solitons exhibit energy exchanging inelastic collision characterized with an intensity redistribution, which has not been found in the single-component cases and may be applied to realize multi-state logic and soliton collision-based computing.
In this study, the multi-component generalization of the Maccari system \cite{maccari}
\begin{align}
& \textmd{i}\phi_t+\phi_{xx}+u \phi=0,\label{j1}\\
& \textmd{i}\psi_t+\psi_{xx}+u \psi=0,\label{jj1}\\
&u_y= (\phi{\phi}^\ast +\psi{\psi}^\ast )_{x},\label{j2}
\end{align}
is considered. In which, $\phi$ and $\psi$ are the complex short-wave amplitudes and $u$ is the real long-wave amplitude; the subscripts denote partial differentiation and the asterisk means complex conjugate hereafter. This system is firstly introduced by Maccari and can be used to describe the motion of isolated waves, localized in small part of space, in a variety of fields such as nonlinear optics, plasma physics and hydrodynamics. It is not difficult to find the relationship of Eqs.(\ref{j1})-(\ref{j2}) with some other well-known models.
For instance, the reduction $y=x$ leads to the celebrated NLS equation \cite{zakharov};
it reduces to the coupled long-wave resonance system \cite{wave} when $y=t$;
and it becomes to the so-called simplest (2+1)-dimensional extension of the NLS equation proposed by Fokas \cite{fokas} if $A=B^*$.
The multi-component counterpart of the Maccari system (\ref{j1})-(\ref{j2}) is given by
\begin{align}
& \textmd{i}\phi^{(k)}_t+\phi^{(k)}_{xx}+u \phi^{(k)}=0,\ \ \ \ \ \ k=1,2,\cdots, M,\label{j3}\\
&u_y=\Big(\sum^M_{k=1} \sigma_k\phi^{(k)}{\phi^{(k)}}^\ast \Big)_{x},\ \ \ \ \ \ \sigma_k=\pm 1,\label{j4}
\end{align}
which describes the interaction of a long wave $u$ with multiple (say $M$) short wave packets $\Phi^{(k)}$.
For convenience, Eqs.(\ref{j3})-(\ref{j4}) are referred to the $M$-component Maccari system hereafter.

For the Maccari system (\ref{j1})-(\ref{j2}), many studies have been done. Uthayakumar etc \cite{k-india} investigate its integrability property by means of the singularity structure analysis.
Lai and Chow \cite{chow} have obtained its two-dromion solutions based on the coalescence of wavenumbers technique. By virtue of the variable separation approach \cite{lou-jpa-1996,lou-jpa-2001,tang}, Zhang etc construct many coherent soliton structures such as dromions, breathers, foldon and solitoff \cite{k-jiefang1,k-jiefang2}.
In a very recent work, its various rational solutions are obtained by Yuan etc \cite{he} through the Hirota's bilinear method.
However, as far as we know, the bright-dark mixed multi-soliton solution to the multi-component Maccari system (\ref{j3})-(\ref{j4}) has not been reported at present.
The purpose of the current paper is to study the bright-dark mixed solitons in the multi-component Maccari system (\ref{j3})-(\ref{j4}) by using the KP hierarchy reduction method.
A general formula of the soliton solution which unifies the all-bright, all-dark and bright-dark mixed multi-soliton solutions of the multi-component Maccari system with all possible combinations of nonlinearities including all-focusing, all-defocusing and mixed types is obtained.
Particularly, dynamical behaviors of the one and two solitons are investigated in detail.
For the collision of two solitons in a $M$-component Maccari system with $M \geq 3$, asymptotic analysis shows that energy-exchanging inelastic collision can take place in the bright parts of the mixed solitons only if the bright parts appear at least in two short-wave components. Moreover, the inelastic collision characterized with an intensity redistribution among the bright parts of the mixed solitons, which can be clearly observed in the collision figures.
While the dark parts of the mixed solitons and the solitons in the long-wave component always undergo elastic collision which just accompanied by a position shift.

The KP hierarchy reduction method for deriving soliton solutions to soliton equations is an elegant and effective technique, which is
firstly introduced by the Kyoto school \cite{jimbo1}. So far, this method has been applied to construct soliton solutions in many equations such as the NLS equation, the coupled higher-order NLS equations, the modified KdV equation and the Davey-Stewartson (DS) equation. Particularly, in Ref. \cite{will1}, the reduction of the two-dimensional Toda lattice hierarchy with constrained KP systems to derive dark solitons is established; and in Ref. \cite{will2}, the reduction of constrained KP systems to get bright solitons from multi-component KP hierarchy is introduced. By means of this method, Ohta et al \cite{ohta} have obtained the general $N$-dark-dark solitons in a two-coupled focusing-defocusing NLS equations (Manakov system). Also based on this method, the mixed $N$-soliton solution to a vector NLS equations is obtained by Feng \cite{feng}. In some other recent works, we have applied this method to study the solitons in the multi-component Yajima-Oikawa (YO) system \cite{chen1,chen2} and the multi-component Mel'nikov system \cite{han1}.

The rest of the paper is as follows. In section 2, the two-bright-one-dark (2-b-1-d) mixed multi-soliton solution of the three-component Maccari system is constructed; moreover, the dynamics of one and two solitons are also investigated in detail.
The one-bright-two-dark (1-b-2-d) mixed multi-soliton solution of the three-component Maccari system and the corresponding dynamics are discussed in section 3.
The similar analysis is extended to get the $m$-bright-($M-m$)-dark mixed multi-soliton solution to the $M$-component Maccari system in section 4. The last section is allotted for conclusion.

\section{2-b-1-d Mixed Multi-soliton Solution of the Three-component Maccari System}
We first consider the 2-b-1-d mixed multi-soliton solution to the three-component Maccari system
\begin{align}
& \textmd{i}\phi^{(k)}_t+\phi^{(k)}_{xx}+u \phi^{(k)}=0,\ \ \ \ k=1,2,3,\label{j5}\\
&u_y=\Big(\sum^3_{k=1} \sigma_k\phi^{(k)}{\phi^{(k)}}^\ast \Big)_{x},\label{j8}
\end{align}
where $\sigma_k=\pm 1$ for $k=1,2,3$. For the three-component Maccari system (\ref{j5})-(\ref{j8}), the mixed-type vector solitons in short-wave components consist of only two types: 2-b-1-d and 1-b-2-d. The 1-b-2-d mixed multi-soliton solution will be derived in the next section.

Without loss of generality, we assume the $\phi^{(1)}$ and $\phi^{(2)}$ components are of bright type and the $\phi^{(3)}$ component is of dark type. The following dependent variable transformations are introduced
\begin{align}\label{jj7}
& \phi^{(1)}= \frac{g^{(1)}}{f},\ \ \ \ \phi^{(2)}= \frac{g^{(2)}}{f},\ \ \ \ \phi^{(3)}=\rho_1 {\rm e}^{\textmd{i}(\alpha_1x-\alpha^2_1t)} \frac{h^{(1)}}{f},\ \ \ \ u=2 (\log f)_{xx},
\end{align}
where $g^{(1)}, g^{(2)}$ and $h^{(1)}$ are complex; $f$ is real; $\alpha_1$ and $\rho_1$ are real constants. Thus the three-component Maccari system (\ref{j5})-(\ref{j8}) is transformed into the bilinear equations
\begin{align}
& (D^2_x+{\rm i} D_t)g^{(k)} \cdot f=0,\ \ \ \ k=1,2,\label{j10}\\
& (D^2_x+2 \textmd{i} \alpha_1 D_x+{\rm i} D_t)h^{(1)} \cdot f=0,\label{j11}\\
& D_xD_yf\cdot f=\sum^2_{k=1}\sigma_k g^{(k)}{g^{(k)}}^\ast-\sigma_3 \rho^2_1 (f^2-h^{(1)}{h^{(1)}}^\ast),\label{j13}
\end{align}
where the Hirota's bilinear operator $D$ is defined as
\begin{align}
& D^l_xD^m_yD^n_tf(x,y,t)\cdot g(x,y,t)=\Big(\frac{\partial}{\partial x}-\frac{\partial}{\partial x'}\Big)^l\Big(\frac{\partial}{\partial y}-\frac{\partial}{\partial y'}\Big)^m\Big(\frac{\partial}{\partial t}-\frac{\partial}{\partial t'}\Big)^nf(x,y,t)\cdot g(x',y',t')|_{x=x',y=y',t=t'}.
\end{align}

In what follows, we proceed to derive the 2-b-1-d mixed multi-soliton solution through the KP hierarchy reduction method. For this purpose, we consider the three-component KP hierarchy with one copy of shifted singular point ($c_1$)
\begin{align}
& (D^2_{x_1}-D_{x_2})\tau_{1,0}(k_1) \cdot \tau_{0,0}(k_1)=0,\label{j20}\\
& (D^2_{x_1}-D_{x_2})\tau_{0,1}(k_1) \cdot \tau_{0,0}(k_1)=0,\label{j21}\\
& (D^2_{x_1}-D_{x_2}+2c_1D_{x_1})\tau_{0,0}(k_1+1) \cdot \tau_{0,0}(k_1)=0,\label{j22}\\
& D_{x_1}D_{y^{(1)}_1}\tau_{0,0}(k_1) \cdot \tau_{0,0}(k_1)=-2\tau_{1,0}(k_1)  \tau_{-1,0}(k_1),\label{j24}\\
& D_{x_1}D_{y^{(2)}_1}\tau_{0,0}(k_1) \cdot \tau_{0,0}(k_1)=-2\tau_{0,1}(k_1)  \tau_{0,-1}(k_1),\label{j25}\\
& (D_{x_1}D_{x^{(1)}_{-1}}-2)\tau_{0,0}(k_1) \cdot \tau_{0,0}(k_1)=-2\tau_{0,0}(k_1+1)  \tau_{0,0}(k_1-1).\label{j25b}
\end{align}
According to the KP theory by Sato \cite{jimbo1}, the bilinear equations (\ref{j20})-(\ref{j25b}) have the tau function solution in Gram determinant form
\begin{align}
& \tau_{0,0}(k_1)=\left| \begin{array}{ccccc}
A & I  \\
-I & B
\end{array} \right|,\\
& \tau_{1,0}(k_1)=\left| \begin{array}{ccccc}
A & I  & \Omega^{\textmd{T}}\\
-I & B & \mathbf{0}^{\textmd{T}}\\
\mathbf{0} & -\bar{\Psi} & 0
\end{array} \right|,\ \ \ \ \ \
 \tau_{-1,0}(k_1)=\left| \begin{array}{ccccc}
A & I  & \mathbf{0}^{\textmd{T}}\\
-I & B & \Psi^{\textmd{T}}\\
-\bar{\Omega} & \mathbf{0} & 0
\end{array} \right|,\\
& \tau_{0,1}(k_1)=\left| \begin{array}{ccccc}
A & I  & \Omega^{\textmd{T}}\\
-I & B & \mathbf{0}^{\textmd{T}}\\
\mathbf{0} & -\bar{\Upsilon} & 0
\end{array} \right|,\ \ \ \ \ \ \
 \tau_{0,-1}(k_1)=\left| \begin{array}{ccccc}
A & I  & \mathbf{0}^{\textmd{T}}\\
-I & B & \Upsilon^{\textmd{T}}\\
-\bar{\Omega} & \mathbf{0} & 0
\end{array} \right|,
\end{align}
in which $\mathbf{0}$ is an $N$-component zero-row vector; $I$ is an $N \times N$ identity matrix; $A$ and $B$ are two $N \times N$ matrices with the entries
\begin{align*}
& a_{ij}(k_1)=\frac{1}{p_i+\bar{p}_j} \Big(-\frac{p_i-c_1}{\bar{p}_j+c_1}\Big)^{k_1} \textmd{e}^{\xi_i+\bar{\xi}_j},\ \ \ \ \ \ b_{ij}=\frac{1}{q_i+\bar{q}_j}\textmd{e}^{\eta_i+\bar{\eta}_j}+ \frac{1}{r_i+\bar{r}_j}\textmd{e}^{\chi_i+\bar{\chi}_j},
\end{align*}
while $\Omega, \Psi, \Upsilon, \bar{\Omega}, \bar{\Psi}$ and $\bar{\Upsilon}$ are $N$-component row vectors defined as
\begin{align*}
&  \Omega=(\textmd{e}^{\xi_1},\textmd{e}^{\xi_2},\cdots,\textmd{e}^{\xi_N}),\ \ \ \ \ \ \Psi=(\textmd{e}^{\eta_1},\textmd{e}^{\eta_2},\cdots,\textmd{e}^{\eta_N}),\ \ \ \ \ \ \Upsilon=(\textmd{e}^{\chi_1},\textmd{e}^{\chi_2},\cdots,\textmd{e}^{\chi_N}),\\
&  \bar{\Omega}=(\textmd{e}^{\bar{\xi}_1},\textmd{e}^{\bar{\xi}_2},\cdots,\textmd{e}^{\bar{\xi}_N}),\ \ \ \ \ \ \bar{\Psi}=(\textmd{e}^{\bar{\eta}_1},\textmd{e}^{\bar{\eta}_2},\cdots,\textmd{e}^{\bar{\eta}_N}),\ \ \ \ \ \ \bar{\Upsilon}=(\textmd{e}^{\bar{\chi}_1},\textmd{e}^{\bar{\chi}_2},\cdots,\textmd{e}^{\bar{\chi}_N}),
\end{align*}
meanwhile
\begin{align*}
& \xi_i=\frac{1}{p_i-c_1}x^{(1)}_{-1}+p_ix_1+p^2_ix_2+\xi_{i0},\ \ \ \ \ \  \bar{\xi}_j=\frac{1}{\bar{p}_j+c_1}x^{(1)}_{-1}+\bar{p}_jx_1-\bar{p}^2_jx_2+\bar{\xi}_{j0},\\
& \eta_i=q_iy^{(1)}_1+\eta_{i0},\ \ \ \ \ \  \bar{\eta}_j=\bar{q}_jy^{(1)}_1+\bar{\eta}_{j0},\ \ \ \ \ \
 \chi_i=r_iy^{(2)}_1+\chi_{i0},\ \ \ \ \ \  \bar{\chi}_j=\bar{r}_jy^{(2)}_1+\bar{\chi}_{j0},
\end{align*}
where $p_i, \bar{p}_j, q_i, \bar{q}_j, r_i, \bar{r}_j, \xi_{i0}, \bar{\xi}_{j0}, \eta_{i0}, \bar{\eta}_{j0}, \chi_{i0}, \bar{\chi}_{j0}$ and $c_1$ are complex parameters.

The bilinear equations (\ref{j20})-(\ref{j25b}) can be proved by using the Grammian technique \cite{hirota}, whose details are omitted here.
Firstly, let's consider complex conjugate reduction by assuming $x_1$, $x^{(1)}_{-1}$, $y^{(1)}_1$ and $y^{(2)}_1$ are real; $x_2$ and $c_1$ are pure imaginary; moreover, taking $p^*_j=\bar{p}_j,q^*_j=\bar{q}_j,r^*_j=\bar{r}_j,\xi^*_{j0}=\bar{\xi}_{j0},\eta^*_{j0}=\bar{\eta}_{j0}$ and $\chi^*_{j0}=\bar{\chi}_{j0}$, then one can find that
\begin{align*}
a_{ij}(k_1)=a^*_{ji}(k_1),\ \ \ \ b_{ij}=b^*_{ji}.
\end{align*}
Furthermore, by defining
\begin{align*}
f=\tau_{0,0}(0),\ \ \ \ g^{(1)}=\tau_{1,0}(0),\ \ \ \ g^{(2)}=\tau_{0,1}(0),\ \ \ \ h^{(1)}=\tau_{0,0}(1),
\end{align*}
it is easy to check $f$ is real and
\begin{align*}
g^{(1)*}=-\tau_{-1,0}(0),\  \ \ \ g^{(2)*}=-\tau_{0,-1}(0),\  \ \ \ h^{(1)*}=\tau_{0,0}(-1),
\end{align*}
hence the equations (\ref{j20})-(\ref{j25b}) become to
\begin{align}
& (D^2_{x_1}-D_{x_2})g^{(k)} \cdot f=0,\ \ \ \ k=1,2,\label{j26}\\
& (D^2_{x_1}-D_{x_2}+2c_1D_{x_1})h^{(1)} \cdot f=0,\label{j27}\\
& D_{x_1}D_{y^{(k)}_1}f \cdot f=2g^{(k)}g^{(k)*},\ \ \ \ k=1,2,\label{j30}\\
& (D_{x_1}D_{x^{(1)}_{-1}}-2)f \cdot f=-2h^{(1)} h^{(1)*}\label{j31}.
\end{align}

Applying the independent variable transformations
\begin{align}
& x_1=x,\ \ \ \ x_2=\textmd{i}(y+t),\label{j42}
\end{align}
equations (\ref{j26})-(\ref{j27}) become to equations (\ref{j10})-(\ref{j11}) with $c_1={\rm i}\alpha_1$.
So the remaining task is to derive equation (\ref{j13}) from equations (\ref{j30})-(\ref{j31}).

What's more, by row operations, $f$ can be rewritten as
\begin{align}
& f=\left| \begin{array}{ccccc}
A' & I  \\
-I & B'
\end{array} \right|,
\end{align}
where the entries in the $N \times N$ matrices $A'$ and $B'$ are given by
\begin{align*}
& a'_{ij}=\frac{1}{p_i+p^*_j},\ \ \ \ \ \
 b'_{ij}=\frac{1}{q_i+q^*_j}\textmd{e}^{\eta_i+\eta^*_j+\xi^*_i+\xi_j}+ \frac{1}{r_i+r^*_j}\textmd{e}^{\chi_i+\chi^*_j+\xi^*_i+\xi_j},
\end{align*}
with
\begin{align*}
& \eta_i+\xi^*_i=q_iy^{(1)}_1+\frac{1}{p^*_i+c_1}x^{(1)}_{-1}+p^*_ix_1-{p_i^*}^2x_2+\xi^*_{i0}+\eta_{i0},\\
& \eta^*_j+\xi_j=q^*_jy^{(1)}_1+\frac{1}{p_j-c_1}x^{(1)}_{-1}+p_jx_1+p_j^2x_2+\xi_{j0}+\eta^*_{j0},\\
& \chi_i+\xi^*_i=r_iy^{(2)}_1+\frac{1}{p^*_i+c_1}x^{(1)}_{-1}+p^*_ix_1-{p_i^*}^2x_2+\xi^*_{i0}+\chi_{i0},\\
& \chi^*_j+\xi_j=r^*_jy^{(2)}_1+\frac{1}{p_j-c_1}x^{(1)}_{-1}+p_jx_1+p_j^2x_2+\xi_{j0}+\chi^*_{j0}.
\end{align*}

Note that, under the reduction conditions
\begin{align}
& -2\textmd{i}{p_i^*}^2=\sigma_1q_i-\frac{\sigma_3\rho^2_1}{p_i^*+c_1},\ \ \ \ \ \ \  2\textmd{i}p^2_j=\sigma_1q^*_j-\frac{\sigma_3\rho^2_1}{p_j-c_1},\\
& -2\textmd{i}{p_i^*}^2=\sigma_2r_i-\frac{\sigma_3\rho^2_1}{p_i^*+c_1},\ \ \ \ \ \ \  2\textmd{i}p^2_j=\sigma_2r^*_j-\frac{\sigma_3\rho^2_1}{p_j-c_1},
\end{align}
i.e.,
\begin{align}
& \frac{1}{q_i+q^*_j}=\frac{\sigma_1}{2\textmd{i}(-{p_i^*}^2+p^2_j)+\frac{\sigma_3\rho^2_1(p_i^*+p_j)}{(p_i^*+\textmd{i}\alpha_1)(p_j-\textmd{i}\alpha_1)}},\\
& \frac{1}{r_i+r^*_j}=\frac{\sigma_2}{2\textmd{i}(-{p_i^*}^2+p^2_j)+\frac{\sigma_3\rho^2_1(p_i^*+p_j)}{(p_i^*+\textmd{i}\alpha_1)(p_j-\textmd{i}\alpha_1)}},
\end{align}
the following relation holds
\begin{align}
& 2\textmd{i}\partial_{x_2}b'_{ij}=(\sigma_1\partial_{y^{(1)}_1}+\sigma_2\partial_{y^{(2)}_1}-\sigma_3\rho^2_1\partial_{x^{(1)}_{-1}})b'_{ij}.\label{n1}
\end{align}
Equation (\ref{n1}) immediately gives
\begin{align}
& 2\textmd{i}f_{x_2}=\sigma_1f_{y^{(1)}_1}+\sigma_2f_{y^{(2)}_1}-\sigma_3\rho^2_1f_{x^{(1)}_{-1}}.\label{j38}
\end{align}
Differentiating (\ref{j38}) with respect to $x_1$, we can get
\begin{align}
& 2\textmd{i}f_{x_1x_2}=\sigma_1f_{x_1y^{(1)}_1}+\sigma_2f_{x_1y^{(2)}_1}-\sigma_3\rho^2_1f_{x_1x^{(1)}_{-1}}.\label{j39}
\end{align}

On the other hand, equations (\ref{j30}) and (\ref{j31}) can be expanded as
\begin{align}
& f_{x_1y^{(1)}_1}f-f_{x_1}f_{y^{(1)}_{1}}=g^{(1)}g^{(1)*},\ \ \ \ \ \ \ \ \ f_{x_1y^{(2)}_1}f-f_{x_1}f_{y^{(2)}_{1}}=g^{(2)}g^{(2)*},\label{j40}
\end{align}
and
\begin{align}
& f_{x_1x^{(1)}_{-1}}f-f_{x_1}f_{x^{(1)}_{-1}}-f^2=-h^{(1)} h^{(1)*},\label{j41}
\end{align}
respectively. By referring to the relations (\ref{j38}) and (\ref{j39}), from (\ref{j40}) and (\ref{j41}), it is easy to get
\begin{align}\label{j41n}
& 2\textmd{i}(f_{x_1x_2}f-f_{x_1}f_{x_2})=-\sigma_1g^{(1)}g^{(1)*}-\sigma_2g^{(2)}g^{(2)*}+\sigma_3\rho^2_1(f^2-h^{(1)}h^{(1)*}),
\end{align}
which is nothing but the equation (\ref{j13}) after applying the transformations (\ref{j42}).

It is worth noting that the variables $y^{(1)}_1, y^{(2)}_1,x^{(1)}_{-1}$ become dummy variables under the transformations (\ref{j42}), hence they can be treated as constants. Consequently, we can let $ {\rm e}^{\eta_{i}}=c^{(1)*}_i$, ${\rm e}^{\eta^*_{i}}=c^{(1)}_i$, $ {\rm e}^{\chi_{i}}=c^{(2)*}_i$, ${\rm e}^{\chi^*_{i}}=c^{(2)}_i$ for $i=1,2,\cdots,N$ and define $ C_k=-(c^{(k)}_1,c^{(k)}_2,\cdots,c^{(k)}_N)$.
Finally, the 2-b-1-d mixed multi-soliton solution of the three-component Maccari system (\ref{j5})-(\ref{j8}) is obtained
\begin{align}\label{j44}
& f=\left| \begin{array}{ccccc}
A & I  \\
-I & B
\end{array} \right|,\ \ \ \ \ \ \
 g^{(k)}=\left| \begin{array}{ccccc}
A & I  & \Omega^{\textmd{T}}\\
-I & B & \mathbf{0}^{\textmd{T}}\\
\mathbf{0} & C_k & 0
\end{array} \right|,\ \ \ \ \ \ \
h^{(1)}=\left| \begin{array}{ccccc}
A^{(1)} & I  \\
-I & B
\end{array} \right|,
\end{align}
where the entries in $A,A^{(1)}$ and $B$ are given by
\begin{align}\label{j45}
& a_{ij}=\frac{1}{p_i+p^*_j} \textmd{e}^{\xi_i+\xi^*_j},\ \ \ \ \ \ a^{(1)}_{ij}=\frac{1}{p_i+p^*_j} \Big(-\frac{p_i-\textmd{i}\alpha_1}{p^*_j+\textmd{i}\alpha_1}\Big) \textmd{e}^{\xi_i+\xi^*_j},\\
& b_{ij}=\Big(\sum_{k=1}^{2}\sigma_kc_i^{(k)*}c_j^{(k)}\Big)\Big[2\textmd{i}(-{p_i^*}^2+p^2_j)+\frac{\sigma_3\rho^2_1(p_i^*+p_j)}{(p_i^*+\textmd{i}\alpha_1)(p_j-\textmd{i}\alpha_1)}\Big]^{-1},
\end{align}
respectively; $\Omega$ and $C_k$ are $N$-component row vectors
\begin{align}
& \Omega=(\textmd{e}^{\xi_1},\textmd{e}^{\xi_2},\cdots,\textmd{e}^{\xi_N}),\ \ \ \ \ \ C_k=-(c^{(k)}_1,c^{(k)}_2,\cdots,c^{(k)}_N);
\end{align}
meanwhile $ \xi_i=p_ix +\textmd{i}p^2_i(y+t)+ \xi_{i0}$; $p_i$, $\xi_{i0}$ and $c^{(k)}_i$, $(k=1,2;i=1,2,\cdots,N)$ are complex constants.

\subsection{One-soliton solution}
To get one-soliton solution, we take $N=1$ in the formula (\ref{j44}), and the corresponding tau functions can be expressed in the form
\begin{align}
& f=1+E_{11^*}\textmd{e}^{\xi_1+\xi^*_1},\\
& g^{(k)}=c_1^{(k)}\textmd{e}^{\xi_1},\ \ \ \ \ k=1,2,\\
& h^{(1)}=1+F_{11^*}\textmd{e}^{\xi_1+\xi^*_1},
\end{align}
where
\begin{align*}
&E_{11^*}=\Big(\sum^2_{k=1}\sigma_kc_1^{(k)}c_1^{(k)*}\Big)\Big[2\textmd{i}(p_1+p_1^*)(p_1^2-{p_1^*}^2)+\frac{\sigma_3\rho^2_1(p_1+p_1^*)^2}{(p_1^*+\textmd{i}\alpha_1)(p_1-\textmd{i}\alpha_1)}\Big]^{-1},\\
& F_{11^*}=-\frac{p_1-\textmd{i}\alpha_1}{p^*_1+\textmd{i}\alpha_1}E_{11^*}.
\end{align*}
Note that, this solution is nonsingular only when $E_{11^*}>0$.

Furthermore, the 2-b-1-d mixed one-soliton solution can be rewritten as
\begin{align}
& \phi^{(k)}=\frac{c^{(k)}_1}{2}  \textmd{e}^{-\eta_1} \textmd{e}^{\textmd{i}\xi_{1I}}{\rm sech}(\xi_{1R}+\eta_1),\ \ \ \ \ k=1,2,\\
& \phi^{(3)}=\frac{\rho_1}{2} \textmd{e}^{\textmd{i}\theta_{1}} [1+\textmd{e}^{2\textmd{i}\phi_1}+(\textmd{e}^{2\textmd{i}\phi_1}-1)\tanh(\xi_{1R}+\eta_1)],\\
& u=2p^2_{1R}{\rm sech}^2(\xi_{1R}+\eta_1),
\end{align}
where $\textmd{e}^{2 \eta_1}=E_{11^*}$, $\textmd{e}^{2\textmd{i}\phi_1}=-(p_1-\textmd{i}\alpha_1)/(p^*_1+\textmd{i}\alpha_1)$, $\xi_1=\xi_{1R}+{\rm i}\xi_{1I}$, the suffixes $R$ and $I$ denote the real and imaginary parts, respectively. Indeed, the amplitude of the bright soliton in the $\phi^{(k)}$ component is $\frac{|c^{(k)}_1|}{2} {\rm e}^{-\eta_1}$. For the dark soliton in the $\phi^{(3)}$ component, $|\phi^{(3)}|$ approaches $|\rho_1|$ as $x,y \rightarrow \pm \infty$. What's more, the intensity of the dark soliton is $|\rho_1|\cos\phi_1$. And the amplitude of the bright soliton in the $u$ component is $2p^2_{1R}$.
It is interesting that we can tune the intensity of the bright parts of the mixed one-soliton but keep the depth of the dark part unchange since the parameters $c^{(k)}_1$ appear explicitly in the amplitude of the bright parts.
As an example, the mixed one-soliton is displayed in Fig. \ref{mix-fig1} with the nonlinearities $(\sigma_1,\sigma_2,\sigma_3)=(1,-1,1)$ at time $t=0$.
The parameters used are $p_1=1-\frac{1}{2}{\rm i}$, $\alpha_1=\rho_1=1$, $\xi_{10}=y=0$, $c^{(2)}_1=1+\frac{1}{2}{\rm i}$ and (a) $c^{(1)}_1=\frac{1}{2}+\frac{1}{4}{\rm i}$; (b) $c^{(1)}_1=\frac{1}{4}{\rm i}$.
It is obvious that when the parameters $c^{(k)}_1$ take different values, the intensities of the bright solitons in $\phi^{(1)}$ and $\phi^{(2)}$ components alter, but the depth of the dark soliton in $\phi^{(3)}$ component and the intensity of the bright soliton in $u$ component remain unaltered.


\subsection{Two-soliton solution}
Taking $N=2$ in the formula (\ref{j44}), the two-soliton solution is obtained, and the tau functions can be expressed in the form
\begin{align}
& f=1+E_{11^*}\textmd{e}^{\xi_1+\xi^*_1}+E_{12^*}\textmd{e}^{\xi_1+\xi^*_2}+E_{21^*}\textmd{e}^{\xi_2+\xi^*_1}+E_{22^*}\textmd{e}^{\xi_2+\xi^*_2}+E_{121^*2^*}\textmd{e}^{\xi_1+\xi_2+\xi^*_1+\xi^*_2},\label{jj49}\\
& g^{(k)}=c^{(k)}_1\textmd{e}^{\xi_1}+c^{(k)}_2\textmd{e}^{\xi_2}+G^{(k)}_{121^*}\textmd{e}^{\xi_1+\xi_2+\xi^*_1}+G^{(k)}_{122^*}\textmd{e}^{\xi_1+\xi_2+\xi^*_2},\ \ \ \ \ k=1,2,\label{jj50}\\
& h^{(1)}=1+F^{(1)}_{11^*}\textmd{e}^{\xi_1+\xi^*_1}+F^{(1)}_{1,2^*}\textmd{e}^{\xi_1+\xi^*_2}+F^{(1)}_{21^*}\textmd{e}^{\xi_2+\xi^*_1}+F^{(1)}_{22^*}\textmd{e}^{\xi_2+\xi^*_2}+F^{(1)}_{121^*2^*}\textmd{e}^{\xi_1+\xi_2+\xi^*_1+\xi^*_2},\label{jj51}
\end{align}
where
\begin{align*}
& E_{ij^*}=\Big(\sum^2_{k=1}\sigma_kc_i^{(k)}c_j^{(k)*}\Big)\Big[2\textmd{i}(p_i+p_j^*)(p_i^2-{p_j^*}^2)+\frac{\sigma_3\rho^2_1(p_i+p_j^*)^2}{(p_i-\textmd{i}\alpha_1)(p_j^*+\textmd{i}\alpha_1)}\Big]^{-1},\\
& E_{121^*2^*}=|p_1-p_2|^2\Big[\frac{E_{11^*}E_{22^*}}{(p_1+p_2^*)(p_2+p_1^*)}-\frac{E_{12^*}E_{21^*}}{(p_1+p_1^*)(p_2+p_2^*)}\Big],\\
& F^{(1)}_{ij^*}=-\frac{p_i-\textmd{i}\alpha_1}{p^*_j+\textmd{i}\alpha_1} E_{ij^*},\\
& F^{(1)}_{121^*2^*}=\frac{(p_1-\textmd{i}\alpha_1)(p_2-\textmd{i}\alpha_1)}{(p^*_1+\textmd{i}\alpha_1)(p^*_2+\textmd{i}\alpha_1)}E_{121^*2^*},\\
& G^{(k)}_{12i^*}=(p_1-p_2)\Big(\frac{c^{(k)}_1E_{2i^*}}{p_1+p_i^*}-\frac{c^{(k)}_2E_{1i^*}}{p_2+p_i^*}\Big).
\end{align*}

To obtain nonsingular solution, the denominator $f$ must be nonzero. To this end, we rewrite $f$ as
\begin{align}
& f=2\textmd{e}^{\xi_{1R}+\xi_{2R}}[\textmd{e}^{\eta_1+\eta_2} \cosh(\xi_{1R}-\xi_{2R}+\eta_1-\eta_2)+\textmd{e}^{\eta_3} \cosh(\xi_{1R}-\xi_{2R}+\eta_3)+\textmd{e}^{\zeta_{R}} \cos(\xi_{1I}-\xi_{2I}+\zeta_{I})],
\end{align}
where
\begin{align}\label{jjj}
& \textmd{e}^{2\eta_1}=E_{11^*},\ \ \ \ \textmd{e}^{2\eta_2}=E_{22^*},\ \ \ \ \textmd{e}^{2\eta_3}=E_{121^*2^*},\ \ \ \ \textmd{e}^{\zeta_{R}+\textmd{i}\zeta_{I}}=E_{12^*}.
\end{align}
Consider the condition for the existence of one-soliton solution, we conclude that $E_{ii^*}>0,i=1,2$ is a necessary condition
while ${\rm e}^{\eta_1+\eta_2}+{\rm e}^{\eta_3}>{\rm e}^{\zeta_{R}}$ is a sufficient condition to guarantee a
nonsingular two-soliton solution.
By virtue of the tau functions (\ref{jj49})-(\ref{jj51}), we can perform asymptotic analysis of the mixed two-soliton as in Refs. \cite{lak1, lak3}, which can help us to elucidate the understanding of the collision of two solitons. Particularly, the asymptotic analysis of two solitons in the $x$-$y$ plane will be carried out in detail. The similar approach can also be applied to investigate the collision dynamics in other planes.

Based on the transformations (\ref{jj7}) and the tau functions (\ref{jj49})-(\ref{jj51}), the asymptotic forms of two colliding solitons $s_1$ and $s_2$ before and after collision can be deduced.

(a) Before collision ($x,y\rightarrow -\infty$)

Soliton $s_1$
\begin{align*}
& \Phi_{1-}^{(k)} \simeq A^{(k)}_{1-}\textmd{e}^{\textmd{i}\xi_{1I}}{\rm sech}(\xi_{1R}+\eta_1),\ \ \ \ \ k=1,2,\\
& \Phi_{1-}^{(3)} \simeq \frac{\rho_1}{2} \textmd{e}^{\textmd{i}\theta_{1}} [1+\textmd{e}^{2\textmd{i}\phi_1}+(\textmd{e}^{2\textmd{i}\phi_1}-1)\tanh(\xi_{1R}+\eta_1)],\\
& u_{1-} \simeq 2p^2_{1R}{\rm sech}^2(\xi_{1R}+\eta_1).
\end{align*}

Soliton $s_2$
\begin{align*}
& \Phi_{2-}^{(k)} \simeq A^{(k)}_{2-}\textmd{e}^{\textmd{i}\xi_{2I}}{\rm sech}(\xi_{2R}+\eta_3-\eta_1),\ \ \ \ \ k=1,2,\\
& \Phi_{2-}^{(3)} \simeq \frac{\rho_1}{2} \textmd{e}^{\textmd{i}(\theta_1+2\phi_1)}[1+\textmd{e}^{2\textmd{i}\phi_2}+(\textmd{e}^{2\textmd{i}\phi_2}-1)\tanh(\xi_{2R}+\eta_3-\eta_1)],\\
& u_{2-} \simeq 2p^2_{2R}{\rm sech}^2(\xi_{2R}+\eta_3-\eta_1).
\end{align*}

(b) After collision ($x,y\rightarrow +\infty$)

Soliton $s_1$
\begin{align*}
& \Phi_{1+}^{(k)} \simeq A^{(k)}_{1+}\textmd{e}^{\textmd{i}\xi_{1I}}{\rm sech}(\xi_{1R}+\eta_3-\eta_2),\ \ \ \ \ k=1,2,\\
& \Phi_{1+}^{(3)} \simeq \frac{\rho_1}{2} \textmd{e}^{\textmd{i}(\theta_1+2\phi_2)}[1+\textmd{e}^{2\textmd{i}\phi_1}+(\textmd{e}^{2\textmd{i}\phi_1}-1)\tanh(\xi_{1R}+\eta_3-\eta_2)],\\
& u_{1+} \simeq 2p^2_{1R}{\rm sech}^2(\xi_{1R}+\eta_3-\eta_2).
\end{align*}

Soliton $s_2$
\begin{align*}
& \Phi_{2+}^{(k)} \simeq A^{(k)}_{2+}\textmd{e}^{\textmd{i}\xi_{2I}}{\rm sech}(\xi_{2R}+\eta_2),\ \ \ \ \ k=1,2,\\
& \Phi_{2+}^{(3)} \simeq \frac{\rho_1}{2} \textmd{e}^{\textmd{i}\theta_1}[1+\textmd{e}^{2\textmd{i}\phi_2}+(\textmd{e}^{2\textmd{i}\phi_2}-1)\tanh(\xi_{2R}+\eta_2)],\\
& u_{2+} \simeq 2p^2_{2R}{\rm sech}^2(\xi_{2R}+\eta_2).
\end{align*}

In the above expressions, $\textmd{e}^{2\textmd{i}\phi_j}=-(p_j-\textmd{i}\alpha_1)/(p^*_j+\textmd{i}\alpha_1)$, $j=1,2$; $( A^{(1)}_{1-}, A^{(1)}_{2-})$ are the amplitudes of the bright parts in the mixed two solitons before collision; $( A^{(1)}_{1+}, A^{(1)}_{2+})$ are the corresponding amplitudes after collision. Where the superscript (subscript) of $A$ denotes the component (soliton) number and $-\ (+)$ represents the soliton before (after) collision. In addition, the various amplitudes are given by
\begin{align*}
& A^{(k)}_{1-}=\frac{c^{(k)}_1}{2\sqrt{E_{11^*}}},\ \ \ \ \  A^{(k)}_{2-}=\frac{G^{(k)}_{121^*}}{2\sqrt{E_{11^*}E_{121^*2^*}}}, \ \ \ \ \ A^{(k)}_{1+}=\frac{G^{(k)}_{122^*}}{2\sqrt{E_{22^*}E_{121^*2^*}}},\ \ \ \ \ A^{(k)}_{2+}=\frac{c^{(k)}_2}{2\sqrt{E_{22^*}}}.
\end{align*}
Furthermore, the amplitudes of the bright parts of the mixed two solitons before and after collision are related through the relation \cite{lak3}
\begin{align}\label{jj53}
& A^{(k)}_{i+}=T_i^{k}A^{(k)}_{i-},\ \ \ \ \ i,k=1,2,
\end{align}
with the transition amplitudes $T_i^{k}$ defined as
\begin{align*}
& T_1^{k}=\Big(\frac{p_1-p_2}{p^*_1-p^*_2}\Big)\Big(\frac{p^*_1+p_2}{p_1+p^*_2}\Big)^{1/2}\Big[\frac{(c^{(k)}_2/c^{(k)}_1)r_1-1}{\sqrt{1-r_1r_2}}\Big],\ \ \
 T_2^{k}=\Big(\frac{p_1-p_2}{p^*_1-p^*_2}\Big)\Big(\frac{p^*_1+p_2}{p_1+p^*_2}\Big)^{1/2}\Big[\frac{\sqrt{1-r_1r_2}}{(c^{(k)}_1/c^{(k)}_2)r_2-1}\Big],\ \ \ k=1,2,
\end{align*}
where
\begin{align*}
& r_1=\frac{p_1+p^*_2}{p_2+p^*_2}\frac{E_{12^*}}{E_{22^*}},\ \ \ \ \ r_2=\frac{p_1^*+p_2}{p_1+p^*_1}\frac{E_{21^*}}{E_{11^*}}.
\end{align*}
The relation (\ref{jj53}) indicates that the intensities of the bright parts in the mixed solitons before and after collision differ in general. The transition amplitudes $T_i^{k}$ become unimodular only under the choice $\frac{|c^{(1)}_1|}{|c^{(1)}_2|}=\frac{|c^{(2)}_1|}{|c^{(2)}_2|}$. This means that the bright parts in the mixed solitons exhibit energy-exchanging inelastic collision, and this inelastic collision characterized by an intensity redistribution of the bright parts of the mixed solitons in the $\phi^{(1)}$ and $\phi^{(2)}$ components.
On the other hand, the intensities of the dark parts of the mixed solitons appearing in the $\phi^{(3)}$ component remain unchange after collision. Therefore, the dark parts in the mixed solitons undertake standard elastic collision.
Additionally, the position shift of soliton $s_1$ ($s_2$) is $\Lambda_1=\eta_3-\eta_1-\eta_2$ ($\Lambda_2=-\Lambda_1$), where $\eta_i,i=1,2,3$ are defined in (\ref{jjj}).
Thus, both the bright and dark parts in the mixed two solitons possess the same magnitude position shift but with opposite signs.
The similar asymptotic analysis of the bright solitons in the $u$ component shows that they also undergo elastic collision.
The collision of two solitons in the three-component Maccari system (\ref{j5})-(\ref{j8}) are displayed in Figs.\ref{mix-fig2} and \ref{mix-fig3} with the nonlinearities $(\sigma_1,\sigma_2,\sigma_3)=(-1,1,1)$.
An example of inelastic collision of the bright solitons in the $\phi^{(1)}$ and $\phi^{(2)}$ components is depicted in Fig.\ref{mix-fig2} with the parametric choice $p_1=1-\frac{1}{4}{\rm i}, p_2=\frac{1}{2}-{\rm i}, c^{(1)}_1=3+\frac{1}{2}{\rm i},c^{(1)}_2=\frac{1}{2}+\frac{1}{3}{\rm i}, c^{(2)}_1=\frac{1}{2}+7{\rm i},c^{(2)}_2=8+\frac{1}{2}{\rm i},\rho_1=1, \alpha_1=\frac{1}{2}$
and $ \xi_{10}=\xi_{20}=0$.
Fig.\ref{mix-fig3} represents an example of elastic collision of the bright solitons in the $\phi^{(1)}$ and $\phi^{(2)}$ components,
where the parameters $c^{(1)}_1,c^{(1)}_2, c^{(2)}_1,c^{(2)}_2$ used in Fig.\ref{mix-fig2} are replaced by $ c^{(1)}_1=1,c^{(1)}_2=\frac{1}{3}, c^{(2)}_1=2,c^{(2)}_2=\frac{2}{3}$.
In Fig.\ref{mix-fig2}, (a) shows an energy-sharing collision with complete suppression of the intensity of a soliton after collision in the $\phi^{(1)}$ component; (b) represents the collision of two bright solitons in the $\phi^{(2)}$ component, where the intensity of a soliton is suppressed while the intensity of the other soliton is enhanced. The dynamical mechanism behind such interesting collision is attributed to an intensity redistribution among the short wave components accompanied by finite amplitude-dependent. In addition, the collisions of two solitons displayed in (c) and (d) are all elastic. In Fig.\ref{mix-fig3}, we do not present the plots for the $\phi^{(3)}$ and $u$ components since they exhibit elastic collision in parallel to the (c) and (d) in Fig.\ref{mix-fig2}.

\section{1-b-2-d Mixed Multi-soliton Solution of the Three-component Maccari System}
In this section, assuming the $\phi^{(1)}$ component is of bright type while the $\phi^{(2)}$ and $\phi^{(3)}$ components are of dark type, the dependent variable transformations
\begin{align}
& \phi^{(1)}= \frac{g^{(1)}}{f},\ \ \ \ \phi^{(2)}=\rho_1 {\rm e}^{\textmd{i}(\alpha_1x-\alpha^2_1t)} \frac{h^{(1)}}{f},\ \ \ \ \phi^{(3)}=\rho_2 {\rm e}^{\textmd{i}(\alpha_2x-\alpha^2_2t)} \frac{h^{(2)}}{f},\ \ \ \ u=2 (\log f)_{xx},
\end{align}
convert the three-component Maccari system (\ref{j5})-(\ref{j8}) into the bilinear forms
\begin{align}
& (D^2_x+{\rm i} D_t)g^{(1)} \cdot f=0,\label{j61}\\
& (D^2_x+2 \textmd{i} \alpha_k D_x+{\rm i} D_t)h^{(k)} \cdot f=0,\ \ \ \ \ \ k=1,2,\label{j62}\\
& D_xD_yf\cdot f=\sigma_1 g^{(1)}{g^{(1)}}^\ast-\sum^2_{k=1}\sigma_{k+1} \rho^2_k (f^2-h^{(k)}{h^{(k)}}^\ast),\label{j63}
\end{align}
where $f$ is real; $g^{(1)}, h^{(1)}$ and $h^{(2)}$ are complex; $\alpha_k$ and $\rho_k , (k=1,2)$ are real constants.

To obtain the 1-b-2-d mixed multi-soliton solution, we consider the two-component KP hierarchy with two copies of shifted singular points ($c_1$ and $c_2$)
\begin{align}
& (D^2_{x_1}-D_{x_2})\tau_{1}(k_1,k_2) \cdot \tau_{0}(k_1,k_2)=0,\label{j68}\\
& (D^2_{x_1}-D_{x_2}+2c_1D_{x_1})\tau_{0}(k_1+1,k_2) \cdot \tau_{0}(k_1,k_2)=0,\\
& (D^2_{x_1}-D_{x_2}+2c_2D_{x_1})\tau_{0}(k_1,k_2+1) \cdot \tau_{0}(k_1,k_2)=0,\\
& D_{x_1}D_{y^{(1)}_1}\tau_{0}(k_1,k_2) \cdot \tau_{0}(k_1,k_2)=-2\tau_{1}(k_1,k_2) \tau_{-1}(k_1,k_2),\\
& (D_{x_1}D_{x^{(1)}_{-1}}-2)\tau_{0}(k_1,k_2) \cdot \tau_{0}(k_1,k_2)=-2\tau_{0}(k_1+1,k_2) \tau_{0}(k_1-1,k_2),\\
& (D_{x_1}D_{x^{(2)}_{-1}}-2)\tau_{0}(k_1,k_2) \cdot \tau_{0}(k_1,k_2)=-2\tau_{0}(k_1,k_2+1) \tau_{0}(k_1,k_2-1).\label{j74}
\end{align}
Based on the KP theory by Sato \cite{jimbo1}, the bilinear equations (\ref{j68})-(\ref{j74}) have the tau function solution
\begin{align}
& \tau_{0}(k_1,k_2)=\left| \begin{array}{ccccc}
A & I  \\
-I & B
\end{array} \right|,\\
& \tau_{1}(k_1,k_2)=\left| \begin{array}{ccccc}
A & I  & \Omega^{\textmd{T}}\\
-I & B & \mathbf{0}^{\textmd{T}}\\
\mathbf{0} & -\bar{\Psi} & 0
\end{array} \right|,\ \ \ \ \ \
 \tau_{-1}(k_1,k_2)=\left| \begin{array}{ccccc}
A & I  & \mathbf{0}^{\textmd{T}}\\
-I & B & \Psi^{\textmd{T}}\\
-\bar{\Omega} & \mathbf{0} & 0
\end{array} \right|,
\end{align}
where $\Omega,\Psi,\bar{\Omega},\bar{\Psi}$  are the $N$-component row vectors defined previously; the entries of the $N \times N$ matrices $A$ and $B$ are given by
\begin{align*}
& a_{ij}(k_1,k_2)=\frac{1}{p_i+\bar{p}_j} \Big(-\frac{p_i-c_1}{\bar{p}_j+c_1}\Big)^{k_1} \Big(-\frac{p_i-c_2}{\bar{p}_j+c_2}\Big)^{k_2} \textmd{e}^{\xi_i+\bar{\xi}_j},\ \ \ \ \
 b_{ij}=\frac{1}{q_i+\bar{q}_j}\textmd{e}^{\eta_i+\bar{\eta}_j},
\end{align*}
with
\begin{align*}
& \xi_i=\frac{1}{p_i-c_1}x^{(1)}_{-1}+\frac{1}{p_i-c_2}x^{(2)}_{-1}+p_ix_1+p^2_ix_2+\xi_{i0},\\
& \bar{\xi}_j=\frac{1}{\bar{p}_j+c_1}x^{(1)}_{-1}+\frac{1}{\bar{p}_j+c_2}x^{(2)}_{-1}+\bar{p}_jx_1-\bar{p}^2_jx_2+\bar{\xi}_{j0},\\
& \eta_i=q_iy^{(1)}_1+\eta_{i0},\ \ \ \ \ \ \ \ \ \ \bar{\eta}_j=\bar{q}_jy^{(1)}_1+\bar{\eta}_{j0},
\end{align*}
where $p_i, \bar{p}_j, q_i, \bar{q}_j, \xi_{i0}, \bar{\xi}_{j0}, \eta_{i0}, \bar{\eta}_{j0}, c_1$ and $c_2$ are all complex parameters.

Similarly, we also first consider complex conjugate reduction by assuming $x_1$, $x^{(1)}_{-1}$, $x^{(2)}_{-1}$, $y^{(1)}_1$ are real; $x_2$, $c_1$ and $c_2$ are pure imaginary and letting $p^*_j=\bar{p}_j$, $q^*_j=\bar{q}_j$, $\xi^*_{j0}=\bar{\xi}_{j0}$ and $\eta^*_{j0}=\bar{\eta}_{j0}$.
It is easy to find that
\begin{align*}
a_{ij}(k_1,k_2)=a^*_{ji}(-k_1,-k_2),\ \ \ \ \ \ \ \ b_{ij}=b^*_{ji}.
\end{align*}
Furthermore, by defining
\begin{align*}
f=\tau_{0}(0,0),\ \ \  g^{(1)}=\tau_{1}(0,0),\ \ \  h^{(1)}=\tau_{0}(1,0),\ \ \  h^{(2)}=\tau_{0}(0,1),
\end{align*}
hence, $f$ is real and
\begin{align*}
g^{(1)*}=-\tau_{-1}(0,0),\ \ \  h^{(1)*}=\tau_{0}(-1,0),\ \ \  h^{(2)*}=\tau_{0}(0,-1).
\end{align*}
Therefore, the bilinear equations (\ref{j68})-(\ref{j74}) become to
\begin{align}
& (D^2_{x_1}-D_{x_2})g^{(1)} \cdot f=0,\label{j75}\\
& (D^2_{x_1}-D_{x_2}+2c_kD_{x_1})h^{(k)} \cdot f=0,\ \ \ \ k=1,2,\label{j76}\\
& D_{x_1}D_{y^{(1)}_1}f \cdot f=2g^{(1)}g^{(1)*},\label{j78}\\
& (D_{x_1}D_{x^{(k)}_{-1}}-2)f \cdot f=-2h^{(k)} h^{(k)*},\ \ \ \ \ k=1,2.\label{j79}
\end{align}
Also consider the transformations (\ref{j42}), take $c_1={\rm i}\alpha_1$ and $c_2={\rm i}\alpha_2$, equations (\ref{j75})-(\ref{j76}) are recast into equations (\ref{j61})-(\ref{j62}).
Hence, the remaining task is to derive equation (\ref{j63}) from equations (\ref{j78})-(\ref{j79}).

Note that, under the reduction conditions
\begin{align}
& -2\textmd{i}{p_i^*}^2=\sigma_1q_i-\frac{\sigma_2\rho^2_1}{p_i^*+c_1}-\frac{\sigma_3\rho^2_2}{p_i^*+c_2},\ \ \ \ \ \ \  2\textmd{i}p^2_j=\sigma_1q^*_j-\frac{\sigma_2\rho^2_1}{p_j-c_1}-\frac{\sigma_3\rho^2_2}{p_j-c_2},
\end{align}
i.e.,
\begin{align}
& \frac{1}{q_i+q^*_j}=\frac{\sigma_1}{2\textmd{i}(-{p_i^*}^2+p_j^2)+\frac{\sigma_2\rho^2_1(p_i^*+p_j)}{(p_i^*+c_1)(p_j-c_1)}+\frac{\sigma_3\rho^2_2(p_i^*+p_j)}{(p_i^*+c_2)(p_j-c_2)}},
\end{align}
the following relation holds
\begin{align}
& 2\textmd{i}f_{x_2}=\sigma_1f_{y^{(1)}_1}-\sigma_2\rho^2_1f_{x^{(1)}_{-1}}-\sigma_3\rho^2_2f_{x^{(2)}_{-1}}.\label{j82}
\end{align}
Differentiating (\ref{j82}) with respect to $x_1$, we can get
\begin{align}
& 2\textmd{i}f_{x_1x_2}=\sigma_1f_{x_1y^{(1)}_1}-\sigma_2\rho^2_1f_{x_1x^{(1)}_{-1}}-\sigma_3\rho^2_2f_{x_1x^{(2)}_{-1}}.\label{j83}
\end{align}

On the other hand, equations (\ref{j78}) and (\ref{j79}) can be expanded as
\begin{align}
& f_{x_1y^{(1)}_1}f-f_{x_1}f_{y^{(1)}_{1}}=g^{(1)}g^{(1)*},\label{j84}
\end{align}
and
\begin{align}
& f_{x_1x^{(1)}_{-1}}f-f_{x_1}f_{x^{(1)}_{-1}}-f^2=-h^{(1)} h^{(1)*},\ \ \ \ \ \ \ \ \ f_{x_1x^{(2)}_{-1}}f-f_{x_1}f_{x^{(2)}_{-1}}-f^2=-h^{(2)} h^{(2)*},\label{j85}
\end{align}
respectively. By using the relations (\ref{j82}) and (\ref{j83}), from (\ref{j84}) and (\ref{j85}), we can arrive at
\begin{align}
& 2\textmd{i}(f_{x_1x_2}f-f_{x_1}f_{x_2})=-\sigma_1g^{(1)}g^{(1)*}+\sigma_2\rho^2_1(f^2-h^{(1)}h^{(1)*})+\sigma_3\rho^2_2(f^2-h^{(2)}h^{(2)*}),\label{j86}
\end{align}
which is nothing but the equation (\ref{j63}) after applying the transformations (\ref{j42}).

Finally, we have obtained the 1-b-2-d mixed multi-soliton solution of the three-component Maccari system (\ref{j5})-(\ref{j8})
\begin{align}\label{j87}
& f=\left| \begin{array}{ccccc}
A & I  \\
-I & B
\end{array} \right|,\ \ \ \ \ \ \
 g^{(1)}=\left| \begin{array}{ccccc}
A & I  & \Omega^{\textmd{T}}\\
-I & B & \mathbf{0}^{\textmd{T}}\\
\mathbf{0} & C_1 & 0
\end{array} \right|,\ \ \ \ \ \ \
h^{(k)}=\left| \begin{array}{ccccc}
A^{(k)} & I  \\
-I & B
\end{array} \right|,
\end{align}
where the  entries in $A,A^{(k)}$ and $B$ are given by
\begin{align}
& a_{ij}=\frac{1}{p_i+p^*_j} \textmd{e}^{\xi_i+\xi^*_j},\ \ \ \ \ \  a^{(k)}_{ij}=\frac{1}{p_i+p^*_j} \Big(-\frac{p_i-\textmd{i}\alpha_k}{p^*_j+\textmd{i}\alpha_k}\Big) \textmd{e}^{\xi_i+\xi^*_j},\\
& b_{ij}=\sigma_1c_i^{(1)*}c_j^{(1)}\Big[2\textmd{i}(-{p_i^*}^2+p_j^2)+\sum^2_{l=1}\frac{\sigma_{l+1}\rho^2_l(p_i^*+p_j)}{(p_i^*+\textmd{i}\alpha_l)(p_j-\textmd{i}\alpha_l)}\Big]^{-1},\label{j88}
\end{align}
meanwhile, $\Omega$ and $C_1$ are defined as
\begin{align}
& \Omega=(\textmd{e}^{\xi_1},\textmd{e}^{\xi_2},\cdots,\textmd{e}^{\xi_N}),\ \ \ \ \ \ C_1=-(c^{(1)}_1,c^{(1)}_2,\cdots,c^{(1)}_N),
\end{align}
with $ \xi_i=p_ix +\textmd{i}p^2_i(y+t)+ \xi_{i0}$; $p_i$, $\xi_{i0}$ and $c^{(1)}_i$, $(k=1,2; i=1,2,\cdots,N)$ are complex constants.

\subsection{One-soliton solution}
Take $N=1$ in the formula (\ref{j87}), the one-soliton solution can be obtained.
In this case, the tau functions can be expressed as
\begin{align}
& f=1+E_{11^*}\textmd{e}^{\xi_1+\xi^*_1},\\
& g^{(1)}=c_1^{(1)}\textmd{e}^{\xi_1},\\
& h^{(k)}=1+F_{11^*}\textmd{e}^{\xi_1+\xi^*_1},\ \ \ \ \ k=1,2,
\end{align}
where
\begin{align*}
& E_{11^*}=\sigma_1c_1^{(1)}c_1^{(1)*}\Big[2\textmd{i}(p_1+p_1^*)(p_1^2-{p_1^*}^2)+\sum^2_{l=1}\frac{\sigma_{l+1}\rho^2_l(p_1+p_1^*)^2}{(p_1-\textmd{i}\alpha_l)(p_1^*+\textmd{i}\alpha_l)}\Big]^{-1},\\
& F_{11^*}=-\frac{p_1-\textmd{i}\alpha_k}{p^*_1+\textmd{i}\alpha_k}E_{11^*}.
\end{align*}
It is worth noting that this solution is nonsingular only when $E_{11^*}>0$.

The 1-b-2-d mixed one-soliton solution can be rewritten as
\begin{align}
& \Phi^{(1)}=\frac{c^{(1)}_1}{2} \textmd{e}^{-\eta_1} \textmd{e}^{\textmd{i}\xi_{1I}} {\rm sech}[\xi_{1R}+\eta_1],\\
& \Phi^{(k+1)}=\frac{\rho_k}{2} \textmd{e}^{\textmd{i}\theta_{k}} \{1+\textmd{e}^{2\textmd{i}\phi_k}+(\textmd{e}^{2\textmd{i}\phi_k}-1)\tanh[\xi_{1R}+\eta_1]\},\ \ \ \ \ k=1,2,\\
& u=2p^2_{1R}{\rm sech}^2[\xi_{1R}+\eta_1],
\end{align}
where $\textmd{e}^{2 \eta_1}=E_{11^*}$, $\textmd{e}^{2\textmd{i}\phi_k}=-(p_1-\textmd{i}\alpha_k)/(p^*_1+\textmd{i}\alpha_k)$, $\xi_1=\xi_{1R}+{\rm i}\xi_{1I}$.
The amplitude of the bright soliton in the $\phi^{(1)}$ component is $\frac{|c^{(1)}_1|}{2} {\rm e}^{-\eta_1}$, and the amplitude of the bright soliton in the $u$ component is $2p^2_{1R}$.
For the dark solitons in the $\phi^{(2)}$ and $\phi^{(3)}$ components, $|\phi^{(k+1)}|,  (k=1,2)$ approaches $|\rho_k|$ as $x,y \rightarrow \pm \infty$.
Additionally, the intensity of the dark soliton in the $\phi^{(k+1)}$ component is $|\rho_k|\cos\phi_k$ for $k=1,2$.
What's more, when $\alpha_1$ and $\alpha_2$ take different values, there are two distinct cases:
(i) $\alpha_1=\alpha_2$ and (ii) $\alpha_1\neq\alpha_2$.
In the former case, it is obvious that $\phi_1=\phi_2$, which means that the dark solitons in the $\phi^{(2)}$ and $\phi^{(3)}$ components are proportional to each other.
Hence, it is viewed as degenerate case.
In the latter case, the condition $\phi_1\neq\phi_2$ implies that the dark solitons in these two short wave components
exhibit different degrees of darkness at their centers.
For this case, the $\phi^{(2)}$ and $\phi^{(3)}$ components are not proportional to each other, thus it is considered as non-degenerate case.
The degenerate and non-degenerate cases are depicted in (a) and (b), respectively in Fig.\ref{mix-fig4}. The parameters are chosen as $p_1=1-\frac{1}{2}{\rm i}$, $c^{(1)}_1=1+{\rm i}$, $\rho_1=\sqrt{2}$, $\rho_2=1$, $\xi_{10}=y=0$ and (a) $\alpha_1=\frac{1}{2}$, $\alpha_2=\frac{2}{3}$; (b) $\alpha_1=\alpha_2=-\frac{1}{2}$ with time $t=0$ under the nonlinearities $(\sigma_1,\sigma_2,\sigma_3)=(1,1,-1)$.


\subsection{Two-soliton solution}
Take $N=2$ in the formula (\ref{j87}), we can get the two-soliton solution.
For this case, the tau functions can be expressed in the form
\begin{align}
& f=1+E_{11^*}\textmd{e}^{\xi_1+\xi^*_1}+E_{12^*}\textmd{e}^{\xi_1+\xi^*_2}+E_{21^*}\textmd{e}^{\xi_2+\xi^*_1}+E_{22^*}\textmd{e}^{\xi_2+\xi^*_2}+E_{121^*2^*}\textmd{e}^{\xi_1+\xi_2+\xi^*_1+\xi^*_2},\\
& g^{(1)}=c^{(1)}_1\textmd{e}^{\xi_1}+c^{(1)}_2\textmd{e}^{\xi_2}+G^{(1)}_{121^*}\textmd{e}^{\xi_1+\xi_2+\xi^*_1}+G^{(1)}_{122^*}\textmd{e}^{\xi_1+\xi_2+\xi^*_2},\\
& h^{(k)}=1+F^{(k)}_{11^*}\textmd{e}^{\xi_1+\xi^*_1}+F^{(k)}_{12^*}\textmd{e}^{\xi_1+\xi^*_2}+F^{(k)}_{21^*}\textmd{e}^{\xi_2+\xi^*_1}+F^{(k)}_{22^*}\textmd{e}^{\xi_2+\xi^*_2} +F^{(k)}_{121^*2^*}\textmd{e}^{\xi_1+\xi_2+\xi^*_1+\xi^*_2},\ \ \ \ \ k=1,2,
\end{align}
where
\begin{align*}
& E_{ij^*}=\sigma_1c_i^{(1)}c_j^{(1)*}\Big[2\textmd{i}(p_i+p_j^*)(p_i^2-{p_j^*}^2)+\sum^2_{l=1}\frac{\sigma_{l+1}\rho^2_l(p_i+p_j^*)^2}{(p_i-\textmd{i}\alpha_l)(p_j^*+\textmd{i}\alpha_l)}\Big]^{-1},\\
& F^{(k)}_{ij^*}=-\frac{p_i-\textmd{i}\alpha_k}{p^*_j+\textmd{i}\alpha_k} E_{ij^*},\\
& E_{121^*2^*}=|p_1-p_2|^2\Big[\frac{E_{11^*}E_{22^*}}{(p_1+p_2^*)(p_2+p_1^*)}-\frac{E_{12^*}E_{21^*}}{(p_1+p_1^*)(p_2+p_2^*)}\Big],\\
& F^{(k)}_{121^*2^*}=\frac{(p_1-\textmd{i}\alpha_k)(p_2-\textmd{i}\alpha_k)}{(p^*_1+\textmd{i}\alpha_k)(p^*_2+\textmd{i}\alpha_k)}E_{121^*2^*},\\
& G^{(1)}_{12i^*}=(p_1-p_2)\Big(\frac{c^{(1)}_1E_{2i^*}}{p_1+p_i^*}-\frac{c^{(1)}_2E_{1i^*}}{p_2+p_i^*}\Big).
\end{align*}

For the same reason, the denominator $f$ must be nonzero.
Similarly, we rewrite $f$ as
\begin{align}
& f=2\textmd{e}^{\xi_{1R}+\xi_{2R}}[\textmd{e}^{\zeta_{R}} \cos(\xi_{1I}-\xi_{2I}+\zeta_{I})+\textmd{e}^{\eta_1+\eta_2} \cosh(\xi_{1R}-\xi_{2R}+\eta_1-\eta_2)+\textmd{e}^{\eta_3} \cosh(\xi_{1R}-\xi_{2R}+\eta_3)],
\end{align}
where $\textmd{e}^{2\eta_1}=E_{11^*}, \textmd{e}^{2\eta_2}=E_{22^*},\textmd{e}^{2\eta_3}=E_{121^*2^*}, \textmd{e}^{\zeta_{R}+\textmd{i}\zeta_{I}}=E_{12^*}$. As discussed above, it is easy to know that $E_{ii^*}>0, (i=1,2)$ is a necessary condition and ${\rm e}^{\eta_1+\eta_2}+{\rm e}^{\eta_3}>{\rm e}^{\zeta_{R}}$ is a sufficient condition to guarantee a nonsingular two-soliton solution.

The asymptotic analysis of the collision of two solitons can be carried out as in the previous section, whose details are omitted here. Particularly, the amplitudes of the bright parts in the mixed solitons before and after collision are given by
\begin{align}\label{jj92}
& (A^{(1)}_{1-},A^{(1)}_{2-},A^{(1)}_{1+},A^{(1)}_{2+})=\Big(\frac{c^{(1)}_1}{2\sqrt{E_{11^*}}}, \frac{G^{(1)}_{121^*}}{2\sqrt{E_{11^*}E_{121^*2^*}}}, \frac{G^{(1)}_{122^*}}{2\sqrt{E_{22^*}E_{121^*2^*}}},\frac{c^{(1)}_2}{2\sqrt{E_{22^*}}}\Big).
\end{align}
Substitution of the expressions for various quantities, we can find that the intensities of the bright parts in the mixed two solitons are same before and after collision, i.e.,
$|A^{(1)}_{j-}|=|A^{(1)}_{j+}|$ for $j=1,2$. Obviously, the amplitudes of the dark solitons in the $\phi^{(k+1)}$ component before and after interaction are same and equal to $\rho_k$, which indicates that the intensities of the dark solitons are unchanged during the collision. What's more, it also can be shown that the bright solitons in the $u$ component undertake elastic collision only.
Compared with the 2-b-1-d case, the collision of solitons in the short-wave and long-wave components for the 1-b-2-d case are all elastic, and there is no energy exchange among the different components. It is necessary to point that this feature is consistent with the mixed solitons in the three-component YO system \cite{lak3,chen2}. The 1-b-2-d mixed two solitons are shown in Fig.\ref{mix-fig5} under the nonlinearities $(\sigma_1,\sigma_2,\sigma_3)=(1,1,-1)$, where the parameters are chosen as $p_1=1-\frac{1}{2}{\rm i}, p_2=\frac{1}{2}-\frac{3}{2}{\rm i}, c^{(1)}_1=\frac{1}{2}+{\rm i},c^{(1)}_2=2, \rho_1=\sqrt{2}, \rho_2=\alpha_2=1, \alpha_1=\frac{1}{2}$ and $\xi_{10}=\xi_{20}=0$.
In Fig.\ref{mix-fig5}, the collisions of two bright solitons in the $\phi^{(1)}$ and $u$ components are displayed in (a) and (d), respectively; and the collisions of two dark solitons with different amplitudes in the $\phi^{(2)}$ and $\phi^{(3)}$ components are depicted in (b) and (c), respectively. It is obvious that the solitons in all the components undertake elastic collision accompanied with a position shift but without shape change.




\section{General Mixed Multi-soliton Solution of the Multi-component Maccari System}
In the same spirit as the three-component Maccari system, the general $m$-bright-$M-m$-dark mixed multi-soliton solution of the $M$-component Maccari system (\ref{j3})-(\ref{j4}) can be obtained. For this purpose, we introduce the following dependent variable transformations
\begin{align}
& \phi^{(k)}= \frac{g^{(k)}}{f},\ \ \ \ \phi^{(l)}=\rho_l {\rm e}^{\textmd{i}(\alpha_lx-\alpha^2_lt)} \frac{h^{(l)}}{f},\ \ \ \ u=2 (\log f)_{xx},
\end{align}
which convert equations (\ref{j3})-(\ref{j4}) into
\begin{align}
& (D^2_x+{\rm i} D_t)g^{(k)} \cdot f=0,\ \ \ \ \ \ \ k=1,2,\cdots,m,\label{jj94}\\
& (D^2_x+2 \textmd{i} \alpha_l D_x+{\rm i} D_t)h^{(l)} \cdot f=0,\ \ \ \ \ \ \ l=1,2,\cdots,M-m,\label{jj95}\\
& D_xD_yf\cdot f=\sum^m_{k=1} \sigma_k g^{(k)}{g^{(k)}}^\ast-\sum^{M-m}_{l=1}\sigma_{l+m} \rho^2_l (f^2-h^{(l)}{h^{(l)}}^\ast).\label{jj96}
\end{align}

Similar to the procedure discussed above, one can show that the following tau functions satisfy the bilinear equations (\ref{jj94})-(\ref{jj96}) and thus provide the general mixed multi-soliton solution to the $M$-component Maccari system (\ref{j3})-(\ref{j4})
\begin{align}\label{j105}
& f=\left| \begin{array}{ccccc}
A & I  \\
-I & B
\end{array} \right|,\ \ \ \ \ \ \
 g^{(k)}=\left| \begin{array}{ccccc}
A & I  & \Omega^{\textmd{T}}\\
-I & B & \mathbf{0}^{\textmd{T}}\\
\mathbf{0} & C_k & 0
\end{array} \right|,\ \ \ \ \ \ \
h^{(l)}=\left| \begin{array}{ccccc}
A^{(l)} & I  \\
-I & B
\end{array} \right|,
\end{align}
where $A,A^{(l)}$ and $B$ are $N \times N$ matrices with the elements given by
\begin{align*}
& a_{ij}=\frac{1}{p_i+p^*_j} \textmd{e}^{\xi_i+\xi^*_j},\ \ \ \ \ \ a^{(l)}_{ij}=\frac{1}{p_i+p^*_j} \Big(-\frac{p_i-\textmd{i}\alpha_l}{p^*_j+\textmd{i}\alpha_l}\Big) \textmd{e}^{\xi_i+\xi^*_j},\\
& b_{ij}=\Big(\sum^m_{k=1}\sigma_kc_i^{(k)*}c_j^{(k)}\Big)\Big[2\textmd{i}(-{p_i^*}^2+p_j^2)+\sum^{M-m}_{l=1}\frac{\sigma_{l+m}\rho^2_l(p_i^*+p_j)}{(p_i^*+\textmd{i}\alpha_l)(p_j-\textmd{i}\alpha_l)}\Big]^{-1},
\end{align*}
meanwhile $\Omega$ and $C_k$ are $N$-component row vectors defined as
\begin{align*}
& \Omega=(\textmd{e}^{\xi_1},\textmd{e}^{\xi_2},\cdots,\textmd{e}^{\xi_N}),\ \ \ \ \ \ C_k=-(c^{(k)}_1,c^{(k)}_2,\cdots,c^{(k)}_N),
\end{align*}
where $\xi_i=p_ix +\textmd{i}p^2_i(y+t) + \xi_{i0}$; $p_i$, $\xi_{i0}$ and $c^{(k)}_i$, $(i=1,2,\cdots,N)$ are complex constants.


The solution obtained admits bright-dark mixed multi-soliton solution to the $M$-component Maccari system (\ref{j3})-(\ref{j4}) with all possible combinations of nonlinearities, including all-focusing, all-defocusing and mixed types.
What's more, as discussed in Ref.\cite{lak3}, the arbitrariness of nonlinearities $\sigma_k$ increases the freedom resulting in rich dynamics of mixed soliton.
The similar asymptotic analysis for the collision of two solitons can be performed in the same manner as previous, which are omitted here. For a $M$-component Maccari system (\ref{j3})-(\ref{j4}) with $M \geq 3$, it can be concluded that energy exchanging inelastic collision is possible only if the bright parts of the mixed solitons appear at least in two short wave components. Furthermore, when $\frac{|c^{(1)}_1|}{|c^{(1)}_2|}=\frac{|c^{(2)}_1|}{|c^{(2)}_2|}=\cdots=\frac{|c^{(k)}_1|}{|c^{(k)}_2|},(k=1,2,\cdots,m)$, the bright solitons in the short wave components will take elastic collision; otherwise, they undergo inelastic collision.
However, the dark solitons appearing in the short wave components and the bright solitons appearing in the long wave component always undertake standard elastic collision.

In addition, similar to the vector NLS equations \cite{feng} and the multi-component YO system \cite{chen2}, the expression of the general mixed multi-soliton solution also unifies the all-bright and all-dark multi-soliton solutions. For instance, the all-bright soliton solution can be directly obtained via letting $m=M$ in the mixed multi-soliton sloution. So it supports the same determinant form as the mixed case. Whereas, the expression of the all-dark soliton solution is different from the one of the mixed case. As pointed in Refs.\cite{feng} and \cite{chen2}, it is known that the all-dark multi-soliton solution can alternatively take the same form as (\ref{j105}) except imposing the following constraints on the parameters
\begin{align}
& 4p_{iI}-\sum^{M}_{l=1}\frac{\sigma_{l}\rho^2_l}{|p_i-\textmd{i}\alpha_l|^2}=0,\ \ \ \ \ \ i=1,2,\cdots,N,
\end{align}
and redefining the matrix $B$ to be an identity matrix, i.e., $b_{ij}$ is 1 when $i=j$ and 0 otherwise.

\section{Conclusion}
In this paper, based on the KP hierarchy reduction method, the general bright-dark mixed multi-soliton solution to the multi-component Maccari system with all possible combinations of nonlinearities are obtained. Taking the three-component Maccari system as a concrete example, its two-bright-one-dark (2-b-1-d) and one-bright-two-dark (1-b-2-d) mixed multi-soliton solution are derived in detail.
In the construction of these two types of solution, it is interesting that the derivation of 2-b-1-d mixed multi-soliton solution starts from a (2+1)-component KP hierarchy with one copy of shifted singular point ($c_1$).
In contrast, for the derivation of 1-b-2-d mixed multi-soliton solution, we start from a (1+1)-component KP hierarchy with two copies of shifted singular points ($c_1$ and $c_2$).
Therefore, it is easy to conclude that the number of the components in KP hierarchy matches the number of the short wave components possessing bright solitons
while the number of the copies of shifted singular points coincides with the number of the short wave components possessing dark solitons. This fact also can be referred to the constructions in the Ref.\cite{jimbo1}.
So it is obvious that the $m$-bright-($M-m$)-dark mixed multi-soliton solution to the $M$-component Maccari system can be obtained from the reduction of a $(m+1)$-component KP
hierarchy with $M-m$ copies of shifted singular points.
Additionally, the general mixed solution obtained also unifies the all-bright and the all-dark multi-soliton solutions as special cases.

the dynamics of single and two solitons are also investigated in detail. Particularly, for the soliton collision, it has been found that for a $M$-component Maccari system with $M \geq 3$, if the bright solitons appear at least in two short-wave components, then inelastic collision takes place, which results in energy exchange among the short wave components supporting bright solitons. After the inelastic collision of two solitons, the intensity of a soliton is suppressed while the intensity of the other soliton is enhanced in general, which can be observed in Fig.\ref{mix-fig2} (b). Moreover, an energy-sharing collision with complete suppression of the intensity of a soliton after collision is displayed in Fig.\ref{mix-fig2} (a).
However, the dark solitons appearing in the short wave components and the bright solitons appearing in the long wave component always undertake elastic collision except for a position shift. It is worth noting that this interesting collision process coincides with the one-dimensional and two-dimensional multi-component YO system \cite{lak1,chen2}.

\section{Acknowledgment}
We would like to express our sincere thanks to S.Y. Lou, K. Maruno, J.C. Chen and other members of our discussion group
for their valuable comments and suggestions. The project is supported by the Global Change Research Program
of China (No.2015CB953904), National Natural Science Foundation of China (No.11675054 and 11435005), and
Shanghai Collaborative Innovation Center of Trustworthy Software for Internet of Things (No. ZF1213).

\end{document}